\documentclass{aa}  

\usepackage{graphicx}

\usepackage{soul}
\usepackage{txfonts}
\usepackage{multirow} 
\usepackage{float}
\usepackage{orcidlink}
\usepackage{xcolor}

\hypersetup{
    colorlinks=true,
    linkcolor=blue,
    citecolor=blue,
    filecolor=magenta,      
    urlcolor=cyan,
    }
    
\begin{document} 
 \title{ISOSCELES Project:}
 \subtitle{II. Modelling galactic B-type stars for fast and $\delta-$slow wind regimes}
   \author{N. Machuca \inst{1} \orcidlink{0000-0001-5638-1437}
          \and
          M. Curé \inst{1} \orcidlink{0000-0002-2191-8692}
          \and
          I. Araya \inst{2}
          \and
          R. O. J. Venero \inst{3,4} 
          \and
          L. S. Cidale \inst{3,4} 
          \and
          C. Arcos \inst{1} \orcidlink{0000-0002-4825-4910} 
          \and
          S. Simón-Díaz \inst{5,6} 
          \and
          A. Lobel \inst{7} \orcidlink{0000-0001-5030-019X} 
                    }
   \institute{Instituto de F\'{\i}sica y Astronom\'{\i}a, Facultad de Ciencias, Universidad de Valpara\'{\i}so, Av. Gran Breta\~na 1111, Valpara\'{\i}so, Chile
   \email{natalia.machuca@postgrado.uv.cl}
         \and
         Centro Multidisciplinario de F\'isica, Vicerrector\'ia de Investigaci\'on, Universidad Mayor, 8580745 Santiago, Chile.
         \and
         Departamento de Espectroscop\'{\i}a, Facultad de Ciencias Astron\'omicas y Geof\'{\i}sicas, Universidad Nacional de La Plata, Paseo del Bosque S/N, BF1900FWA La Plata, Buenos Aires, Argentina.
         \and
         Instituto de Astrof\'{\i}sica La Plata, CCT La Plata, CONICET-UNLP, Paseo del Bosque S/N, B1900FWA La Plata, Argentina 
         \and
         Instituto de Astrofísica de Canarias, E-38200 La Laguna, Tenerife, Spain.
         \and
         Departamento de Astrof\'isica, Universidad de La Laguna, E-38205 La Laguna, Tenerife, Spain.
         \and
         Royal Observatory of Belgium, Ringlaan 3, B-1180 Brussels, Belgium.
         }

   \date{Received ; accepted }

  \abstract

\abstract
{Radiation-driven winds in B-type stars play a key role in their evolution, yet their hydrodynamical structure remains uncertain, particularly in evolved objects. While the classical fast solution of the modified CAK (m--CAK) theory is widely adopted, it does not always reproduce the optical wind diagnostics of B-type giants and supergiants.}
{We investigate the applicability of the classical fast and the $\delta$--slow hydrodynamical solutions to B-type stellar winds through a homogeneous spectroscopic analysis based on optical diagnostics.}
{We analysed 50 Galactic B-type spectra spanning luminosity classes~I to~V using mid- and high-resolution optical spectra from the IACOB, ESO-UVES, and CASLEO datasets. Synthetic spectra were taken from the ISOSCELES grid, which combines hydrodynamical wind models computed with \textsc{Hydwind} and NLTE radiative transfer with \textsc{Fastwind}. Stellar and wind parameters were derived via a multi-line $\chi^2$ fitting procedure using hydrogen, helium, and silicon lines.}
{We find evidence for a statistical preference for different hydrodynamical regimes across luminosity classes within the adopted modelling framework. A large fraction of B-type supergiants ($\sim$96\%) and giants ($\sim$88\%) are better reproduced by models based on the $\delta$--slow hydrodynamical solution, characterised by higher ionisation parameters, slower terminal velocities ($v_\infty \lesssim 300$\,km\,s$^{-1}$), and denser outflows. In contrast, most dwarfs and subgiants ($\sim$88\%) in the sample are more frequently consistent with the classical fast solution, with higher $v_\infty$ and lower $\dot{M}$. This behaviour suggests a dichotomy between luminosity classes, further supported by systematic trends of $\delta$, $\dot{M}$, and $v_\infty$ with effective temperature.}
{These results suggest that the $\delta$--slow solution may provide a viable and physically motivated framework for modelling the optical spectra of evolved B-type stars, whereas fast solutions remain adequate for less evolved objects. However, given that optical diagnostics have limited sensitivity to the outer wind regions, these conclusions should be regarded as indicative. By employing hydrodynamically motivated wind structures rather than prescribed velocity laws, the ISOSCELES grid offers a consistent basis for interpreting optical wind diagnostics and motivates future combined multi-wavelength studies to further constrain B-type stellar winds.}
  

   \maketitle
\nolinenumbers
\section{Introduction}

Stellar winds in massive stars constitute one of the most fundamental processes in stellar evolution, being responsible for significant mass loss during their lifetimes and exerting a determining influence on their evolutionary fate. In these stars, winds are driven by the transfer of momentum of photons from the stellar surface to the wind plasma through scattering processes in metal lines  \citep{Castor1975}. This mechanism, which provides the physical basis for the radiation-driven wind, was initially formalised by \citet{lucy1970}, \citet{Castor1974} and \citet{Castor1975} using the Sobolev approximation, thereby establishing the theoretical foundations for the modern understanding of these phenomena.

Subsequent theoretical developments have led to significant improvements known as the modified CAK (m-CAK) theory, which incorporates the effects of rotation and the finite disk correction factor \citep{Pauldrach1986, Friend1986}. In this theoretical formulation, the line radiation force is parametrised by three line-force parameters: $k$, $\alpha$, and $\delta$ \citep{Abbott1982}, which describe the strength and distribution of spectral lines and their sensitivity to ionisation, respectively \citep{Puls2000}. The evolution of this theory has resulted in the identification of three physically distinct types of wind solutions within the m-CAK framework: the standard (fast) solution \citep{Pauldrach1986, Friend1986}, the $\Omega$-slow solution for fast rotators \citep{Cure2004}, and the $\delta$-slow solution for stars with high values of $\delta$, which could apply to B- and A-type supergiants \citep{Cure2011,venero2016}. See more details about these solutions in \citet{cure2023}.

Among these, the $\delta$-slow solution has not yet been widely tested on large stellar samples using synthetic line models, having been only partially explored in previous studies \citep{venero2024, Ortiz2025}. This gap in observational validation represents a significant limitation in our understanding of wind physics in evolved massive stars, preventing robust constraints on hydrodynamical regimes and accurate predictions of their evolution.

In observational stellar wind analyses, most studies of massive stars adopt a very simple velocity-profile approximation: the classical $\beta$-law \citep{Pauldrach1986, Kudritzki2000}. This prescription provides a convenient parametrisation of the wind acceleration and is commonly used to determine two key wind properties: the mass-loss rate ($\dot{M}$) and the terminal velocity ($v_\infty$). Together with the velocity-law exponent ($\beta$), these quantities are typically obtained through quantitative spectroscopy \citep[e.g.][]{simon-diaz2020}, where all three parameters are treated as free variables during the fitting process.

However, several studies have identified significant challenges with these empirical fittings. Several authors \citep[][among others]{Puls2008, Markova2008, Haucke2018, Lefever2023} have noted the need to adopt very high values of $\beta$ ($\geq 2$) to adequately reproduce the H$\alpha$ line profile in B supergiants. This requirement suggests that the standard hydrodynamics of the classical m-CAK theory (fast solution, implying $\beta < 1$) may not adequately describe the wind structure of these evolved objects.

In recent years, several alternative hydrodynamical approaches, independent of the classical m-CAK formalism, have been developed to consistently compute stellar wind structures from first principles. These include methods based on Monte Carlo radiative transfer \citep[e.g.][]{Vink2000, Vink2001}, comoving-frame (CMF) \citep[e.g.][]{sundqvist2019} radiative transfer coupled to hydrodynamics \citep[e.g.][]{Gormaz2019, Bjorklund2021}, and hybrid or iterative schemes combining detailed radiative transfer with dynamical solutions \citep[e.g.][]{Sander2020, BerniniPeron2025}. As summarised in the recent review by \citet{Vink2022}, these complementary approaches have opened new avenues for exploring stellar-wind physics beyond the traditional m-CAK framework, providing more robust and physically consistent predictions for the mass-loss rates and velocity structures of massive stars.

In the frame of quantitative spectroscopy, two primary methodological approaches have emerged for stellar wind analysis. The first consists of computing synthetic spectra from hydrodynamic and stellar-atmosphere models (typically adopting a prescribed velocity field, such as a $\beta$-law) and iteratively adjusting the model parameters until the best match to the observed line profiles is achieved \citep{Mokiem2006, tramper2014, hawcroft2021, brands2022, Lefever2007, SimonDiaz2011, Holgado2022, deBurgos2025}. 

The second approach involves constructing extensive pre-computed grids of stellar atmosphere models and performing an automated or semi-automated search for the model that best reproduces the observed spectra, usually through statistical techniques such as $\chi^2$ minimisation. This strategy enables a more systematic exploration of parameter space and is particularly well suited for the analysis of large spectroscopic samples.

Recent developments have led to the construction of the ISOSCELES grid \citep{araya2025}, which represents a significant advancement in stellar atmosphere and hydrodynamic modelling. This grid comprises hydrodynamic and stellar-atmosphere models of massive stars, designed explicitly for synthetic line-profile analysis. Uniquely, it incorporates both spherical wind solutions (fast and $\delta$-slow) calculated using the Hydwind code \citep{Cure2004}, which serve as input for the non-LTE line-blanketing atmosphere code \textsc{Fastwind} \citep{Puls2005}. This integration provides a flexible and physically consistent framework for analysing large samples of spectra from massive stars, achieving an optimal balance between computational speed and modelling accuracy. We note, however, that the approach is not fully self-consistent, since the line-force parameters are adopted from ISOSCELES' hydrodynamical grid rather than being calculated iteratively from the radiative transfer.

The motivation for this study stems from the need to systematically investigate whether B-type stars consistently exhibit the slower stellar winds predicted by the $\delta$-slow wind regime, particularly in evolved phases. While theoretical predictions suggest this behaviour, comprehensive observational validation using physically consistent models has been lacking. This work aims to address this gap by conducting a systematic analysis of a substantial sample of B-type stars across different evolutionary phases.

Our investigation aims to deepen the understanding of radiation-driven wind behaviour in B-type stars and to emphasise the critical importance of using physically consistent models when interpreting spectral data from these objects. Through this approach, we aim to provide empirical validation of theoretical predictions and to advance the methodological framework for stellar wind analysis in the era of extensive spectroscopic surveys.

The paper is organised as follows: In Section~\ref{sec:obs}, we describe the observational data and instruments used to acquire the sample of 50 B-type stars. Section~\ref{sec:method} details the methodology, including the construction of the ISOSCELES grid, the spectral fitting pipeline, and the diagnostic lines employed in the analysis. Section~\ref{sec:results} presents the results, highlighting representative spectral fits, trends in stellar and wind parameters, and the correlation between wind solutions and luminosity class. Finally, in Section~\ref{sec:dis}, we discuss our results and summarise our main conclusions.

\section{Observations}\label{sec:obs}

Mid- and high-resolution optical spectra of B-type galactic stars were collected from three sources: the IACOB project database \citep{SimonDiaz2011, SimonDiaz2015}, archival data from ESO-UVES, and observations taken from the Complejo Astronómico El Leoncito (CASLEO, Argentina). We analysed a total of 50 B-type spectra spanning different luminosity classes, including 25 supergiants, 8 giants, and 17 dwarf and subgiant stars; see Tables~\ref{tab:stellar_params}.

The sample was selected to represent the B-type spectral domain across a range of effective temperatures and luminosity classes, with an emphasis on evolved stars in which slow winds are expected. To enhance the supergiant regime, we complemented IACOB data with ESO–UVES and CASLEO spectra, combining mid- and high-resolution observations suitable for wind diagnostics.

A total of 32 spectra were retrieved from the IACOB database. These spectra were acquired using either the FIES spectrograph mounted on the $2.56$-m Nordic Optical Telescope (NOT, $R \sim 46\,000$) or the HERMES spectrograph on the $1.2$-m Mercator Telescope ($R \sim 85\,000$), both facilities at the Roque de los Muchachos Observatory, La Palma, Spain. The signal-to-noise (S/N) ratios range from $150$ to $300$. The spectra were retrieved from the IACOB database, already reduced and normalised using the standard pipelines FIEStool \citep{Telting2014} and HERMES-DRS \citep{Raskin2011}, ensuring a homogeneous, high-quality dataset for both early- and evolved-B-type stars.

Nine B-type supergiant stars were observed at the CASLEO using the REOSC spectrograph attached to the $2.15$-m Jorge Sahade (JS) telescope.
The adopted instrumental configuration was a $400$~l/mm grating (\#$580$), a single slit of width $250~\mu$m, and a $1024\times1024$ TEK CCD detector. The spectra have a resolving power of $R \sim 13\,900$ and S/N ratios ranging from $100$ to $300$. Spectra were reduced and wavelength-calibrated using standard IRAF\footnote{IRAF is distributed by the National Optical Astronomy Observatory, which is operated by the Association of Universities for Research in Astronomy (AURA) under cooperative agreement with the National Science Foundation.} routines.

The remaining nine B-supergiants were selected from the ESO Science Archive and correspond to data obtained with the Ultraviolet and Visual Echelle Spectrograph (UVES), mounted at the Nasmyth B focus of the UT2 unit of the Very Large Telescope (VLT) at the Paranal Observatory (Chile). The UVES spectra offer high spectral resolution ($R \sim 80\,000$) and excellent quality, with typical S/N values of around 300. These UVES spectra are available in the online ESO Science Archive. They were retrieved and selected based
on the strength of the H$\alpha$ P Cyg line profile and also the continuum normalised by one of us.
Information about the observational date is provided in the Table \ref{tab:stellar_params}.

The instrumental setup of each subsample provides sufficient wavelength coverage and resolution for modelling both photospheric and wind-sensitive features in these stars. Table~\ref{tab:observations} summarises the characteristics of each subsample, including the instrumentation, spectral resolution, S/N ratio, and number of stars. 

\begin{table}[ht]
\tabcolsep 3pt
\caption{Summary of the B-type star sample analysed in this work. }
\label{tab:observations}
\begin{tabular}{lccc}
\hline\hline
 & IACOB & CASLEO & ESO-UVES \\
\hline
No. of stars & $32$  & $9$  & $9$  \\
Instrument & FIES / HERMES & REOSC & UVES \\
Telescope & NOT / Mercator & JS $2.15$m & VLT UT2 \\
Resolution & $46\,000$ - $85\,000$ & $\sim13\,900$ & $\sim80\,000$ \\
S/N range & $150$ - $300$ & $100$ - $300$ & $\sim300$ \\
\hline
\end{tabular}
\end{table}

\section{Methodology}
\label{sec:method}
\subsection{Model atmospheres and the ISOSCELES grid}

To derive the stellar and wind parameters of our sample, we used a grid-based fitting approach applied to mid- and high-resolution optical spectra. For this purpose, we employed the ISOSCELES grid that combines hydrodynamics and non-LTE radiative transfer covering the $T_\mathrm{eff}$–$\log g$ plane from B dwarf to  B supergiant domain. A unique combination of microturbulence, silicon abundance, solar helium abundance (assumed to be representative of Galactic B-type stars), wind solution type, and terminal velocity characterises each grid point.

This grid spans a wide range of effective temperatures ($9000$~K - $45000$~K), logarithm of surface gravities ($0.6$ - $4.5$), and wind regimes, including both fast and $\delta$-slow m-CAK solutions. 

The construction of the ISOSCELES grid began with the computation of stationary wind solutions using the \textsc{Hydwind} code for different combinations of the line-force parameters $\alpha$, $k$, and $\delta$. For each hydrodynamical solution, the \textsc{Fastwind} code subsequently computed the corresponding synthetic spectra. In \textsc{Fastwind}, the photospheric structure is treated using a spherically symmetric geometry, which is smoothly connected to the hydrodynamical velocity field provided by \textsc{Hydwind} in the wind region. Radiative transfer is solved through the standard iterative scheme of \textsc{Fastwind}, which begins with a Sobolev approximation for the line force during the initial iteration and later switches to a comoving-frame (CMF) treatment, ensuring an accurate description of line formation throughout the photosphere--wind transition (see \citealt{araya2025} for further details).

\subsection{Spectral fitting pipeline}

To perform a semi-automatic quantitative spectroscopic analysis of the B-type stars in our sample, we developed a custom Python-based pipeline for the spectral fitting, designed to be efficient and fully reproducible. The routine compares synthetic spectra from the ISOSCELES grid with observed spectra through a parallelised $\chi^{2}$-minimisation procedure. The $\chi^{2}$ statistic is computed independently for each diagnostic line, and the final value is obtained by averaging the individual $\chi^{2}$ values, giving all lines the same weight. This approach is intended to rank models rather than to provide a formal statistical likelihood, as the uncertainties in the observed spectra are not purely noise-dominated and may include systematic errors.

The diagnostic lines used in the fitting procedure are: H$\alpha$, H$\beta$, H$\gamma$, \ion{He}{i} $\lambda 4471$, \ion{He}{i} $\lambda 6678$ or \ion{He}{ii} $\lambda 4686$, \ion{Si}{ii} $\lambda 4130$, \ion{Si}{iii} $\lambda 4552$, \ion{Si}{iv} $\lambda 4056$ and \ion{Si}{iv} $\lambda 4212$,. Each line responds differently to parameters such as wind density, temperature, gravity, or ionisation, and their combined behaviour constrains both photospheric and wind parameters.

Prior to the comparison between observed and synthetic spectra, a preprocessing step is applied to account for small residual uncertainties in the continuum normalisation of the observed data. This is implemented by applying a uniform vertical scaling (or offset) to the observed spectrum, ensuring that both the observed and synthetic spectra are placed on a consistent continuum level. This correction is applied globally and does not modify the shape of the line profiles, which remain the primary diagnostic for the fitting procedure.

The pipeline operates as follows:

\begin{enumerate}
    \item Input reading: The code parses an input file containing the observed spectra and requires the spectral convolution parameters to account for rotational ($v\sin i$) and macroturbulence ($v_{macro}$) line broadening, typically derived from silicon, oxygen or, in some cases, helium line profiles using the \texttt{iacob\_broad} tool \citep{SimonDiaz2014}. It also optionally accepts preliminary estimates of $T_\mathrm{eff}$, $\log g$, or wind regime (fast or $\delta$-slow) to narrow down the grid search.
    
    \item Parallelised evaluation: Model comparisons are distributed across multiple processors (CPUs) to enhance computational efficiency. Each CPU performs the following tasks:
    \begin{itemize}
        \item Convolves synthetic spectra to match the instrumental resolution, the observed projected rotational velocity, $v\,\sin\,i$, and the macroturbulent broadening, $v_{macro}$.
        \item Interpolates the convolved models onto the observed data's wavelength values.
        \item Computes the $\chi^2$ statistics over selected diagnostic line regions.
        \item Apply a horizontal (wavelength) shift to the observed spectrum to account for the radial velocity of the star.
    \end{itemize}

    \item Model selection and output: Once all models have been evaluated, the results are ordered according to the best-fit models selected based on the lowest $\chi^2$. Final outputs include the derived stellar and wind parameters, diagnostic line plots, and a summary report exported in PDF format. 
\end{enumerate}

Appendix~\ref{fig:code_diagram} provides a schematic diagram of the pipeline architecture. The diagram illustrates how each model is processed in parallel through convolution, interpolation, and $\chi^{2}$ evaluation before the results are combined for final selection.

To illustrate the performance of the ISOSCELES grid and the $\chi^2-$ based fitting procedure, we present six representative stars from our sample in the next section.

\section{Results}\label{sec:results}

Although the present approach relies on prescribed line-force parameters rather than fully self-consistent calculations, it allows us to explore the range of hydrodynamical solutions compatible with the observed optical diagnostics. Therefore, our results should be interpreted as indicative of plausible wind regimes rather than as unique solutions.

It is important to outline the typical uncertainties associated with the derived stellar and wind parameters. Following the methodology described by \citet{araya2025}, the uncertainties in the ISOSCELES fitting procedure are dominated by the resolution of the underlying grid and by the intrinsic degeneracies of the line diagnostics. A quantile-based sensitivity analysis of the best-fitting models (see Appendix B in \citealt{araya2025}, similar to \citet{turis2025}) shows that the typical uncertainties are of the order of $\pm 1000$--$1500$\,K in $T_{\rm eff}$, $\pm 0.15$--$0.25$\,dex in $\log g$, and $\pm 0.2$\,dex in the silicon abundance. For microturbulence, variations of $3$--$5$\,km\,s$^{-1}$ are commonly found. Notice that the stellar radius has been calculated using the flux-weighted gravity-luminosity relationship (FGLR, \citealt{kudritzki2003}), as described in \citet{araya2025}.

The parallelisation strategy adopted in the fitting pipeline is based on a parallel task distribution. Each CPU core processes independent spectral fits, evaluating different grid models for a given star, with no communication required between tasks during the $\chi^{2}$ evaluation. This approach ensures near-linear scaling with the number of available CPUs and allows an efficient exploration of the multi-dimensional ISOSCELES grid, as illustrated in Fig.~\ref{fig:code_diagram}.

Taking into account the methodological aspects and parameter uncertainties discussed above, we now present the results for the stellar sample. A quantitative spectroscopic analysis was performed for 50 B-type stars compiled from three independent sources: IACOB, CASLEO, and ESO-UVES.

\begin{figure*}[ht]
\centering
\includegraphics[width=\hsize]{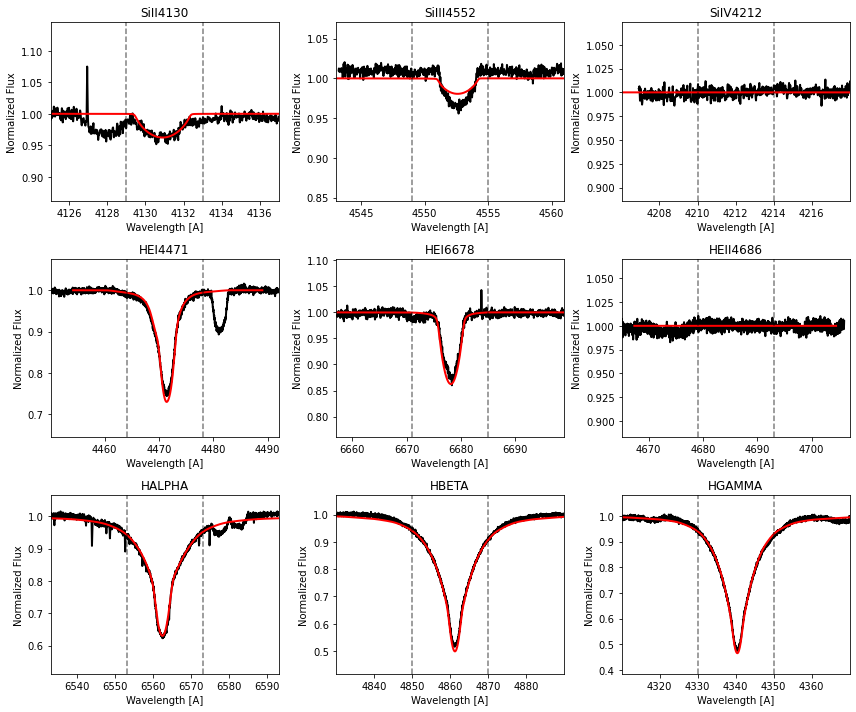}
\caption{Best-fitting synthetic spectrum (solid red line) compared to the observed spectrum (solid black line) of HD~74280 (B4\,V, IACOB). Diagnostic hydrogen, helium, and silicon lines are shown. 
The comparison illustrates agreement between the observations and the fast-solution synthetic model in the line profiles.
}
\label{fig:fit_HD74280}
\end{figure*}

\begin{figure*}[ht]
\centering
\includegraphics[width=\hsize]{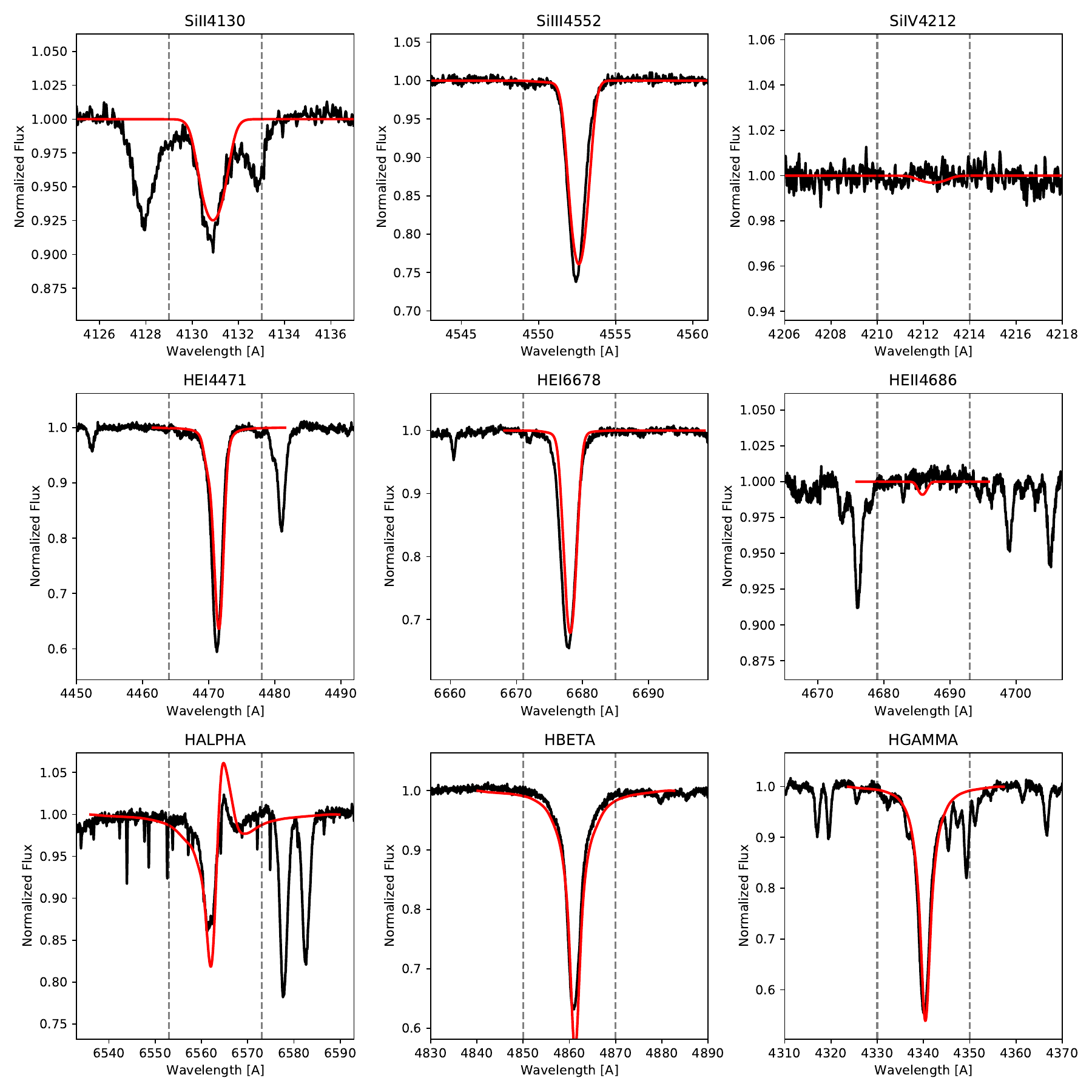}
\caption{Same as Fig.~1, but for HD~206165 (B2\,Ib, IACOB). 
The synthetic profile (solid red line) corresponds to a $\delta$--slow solution. 
Observed spectra are shown with solid black lines, and line styles are the same as in Fig.~\ref{fig:fit_HD74280}.
}
\label{fig:fit_HD206165}
\end{figure*}

\begin{figure}[ht]
\centering
\includegraphics[width=\hsize]{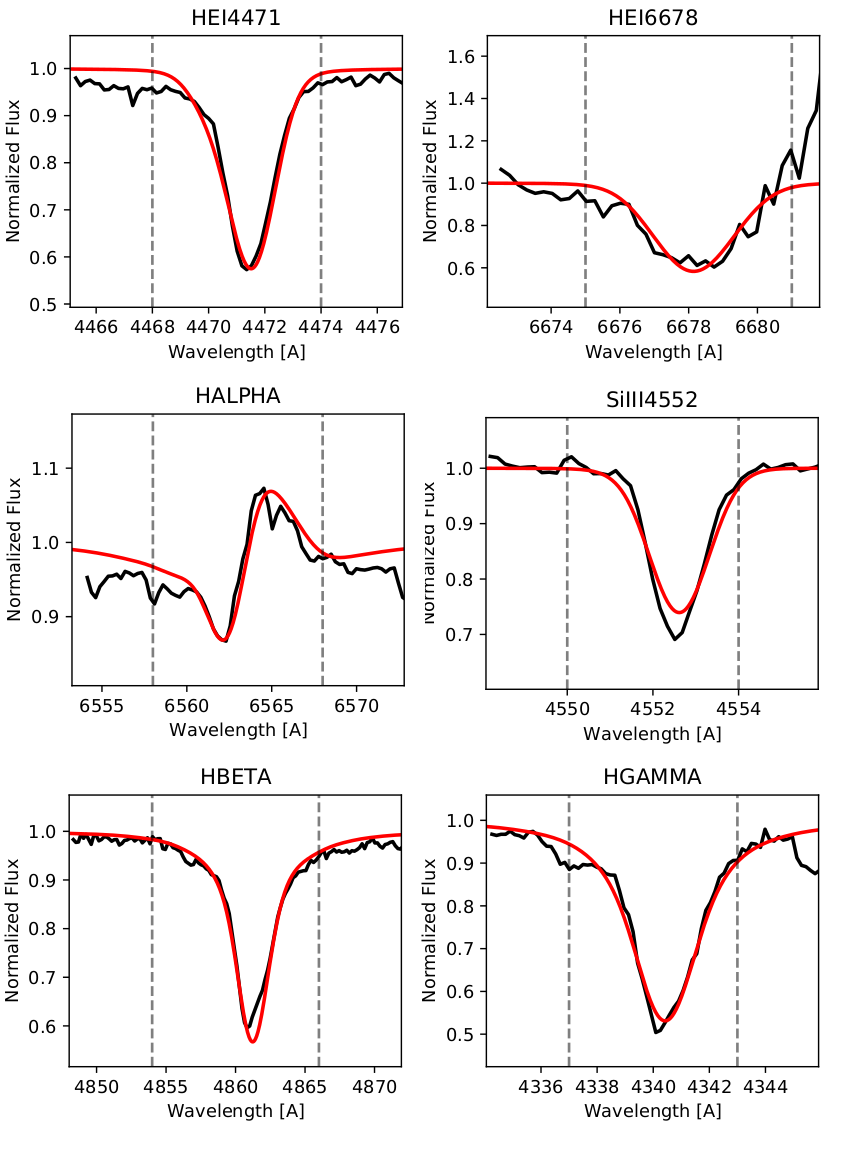}
\caption{Same as Fig.~\ref{fig:fit_HD74280}, but for HD~99953 (B1\,Ib, CASLEO). 
The synthetic spectrum (solid red line) corresponds to a $\delta$--slow wind model. 
Observed CASLEO data are shown as solid black lines.}
\label{fig:fit_HD99953}
\end{figure}

\begin{figure}[ht]
\centering
\includegraphics[width=\hsize]{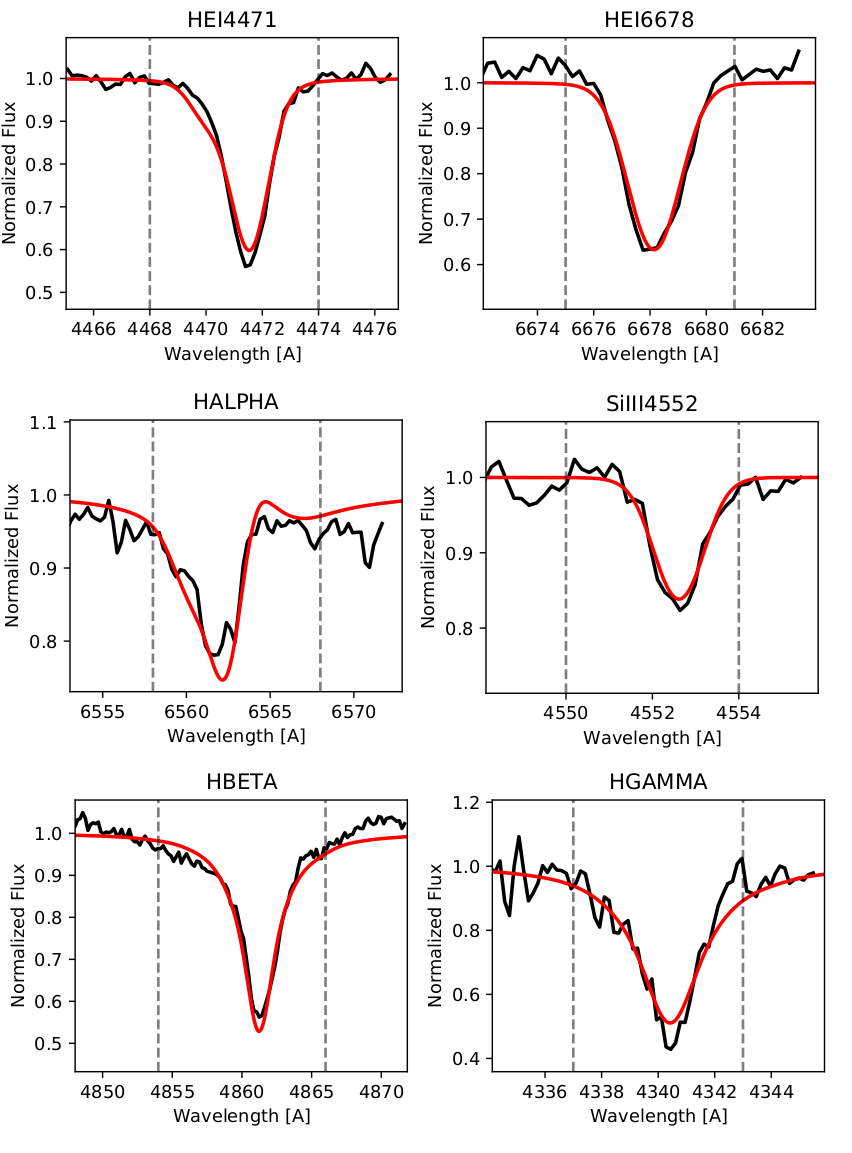}
\caption{Same as Fig.~\ref{fig:fit_HD74280}, but for HD~75149 (B3\,Ia, CASLEO). 
The synthetic $\delta$--slow solution is shown as a solid red line, while the CASLEO observations are plotted in black. Line styles follow the convention used in Fig.~\ref{fig:fit_HD74280}.}
\label{fig:fit_HD75149}
\end{figure}

\begin{figure}[ht]
\centering
   \includegraphics[width=\hsize]{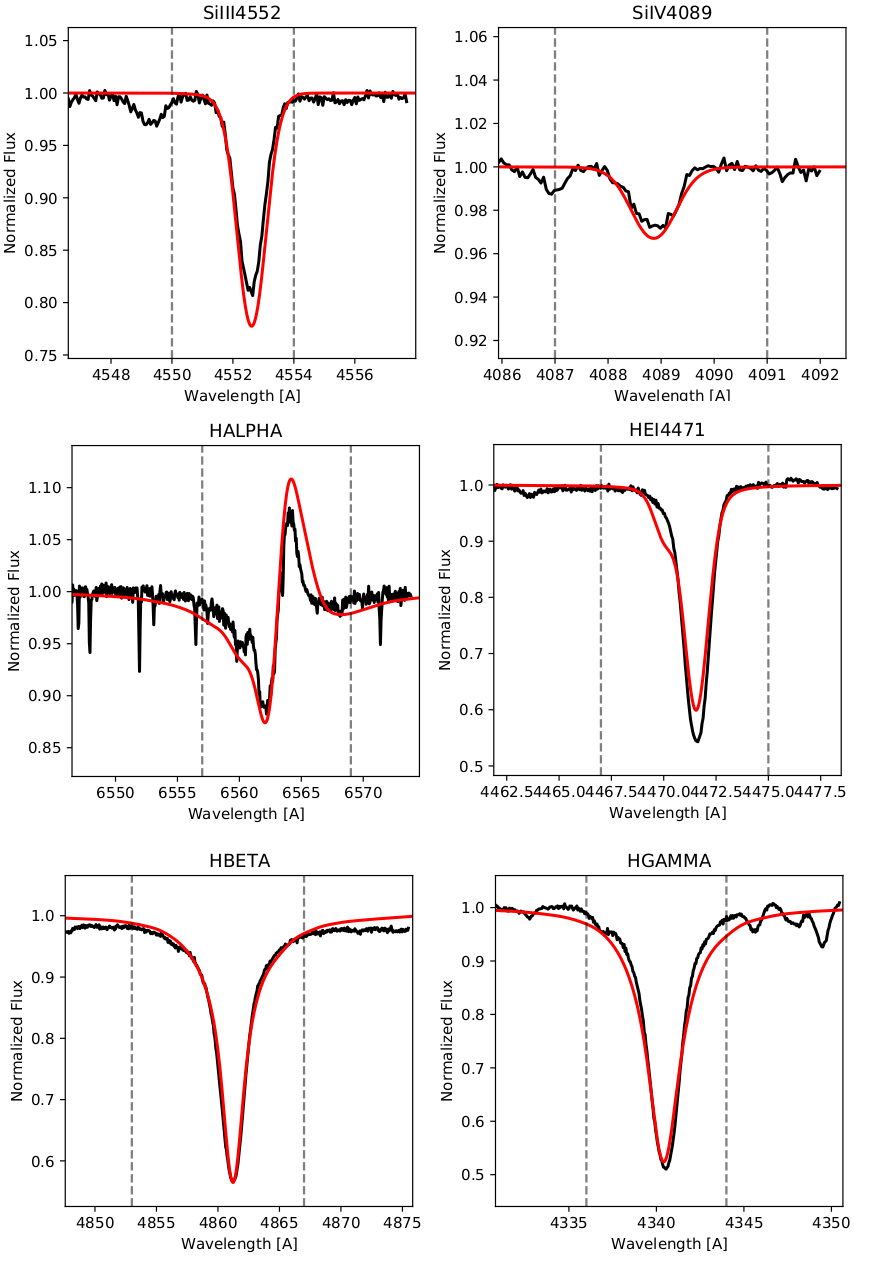}
\caption{Same as Fig.~\ref{fig:fit_HD74280}, but for HD~79186 (B3\,Ib, ESO--UVES). 
The solid red line shows the best-fitting $\delta$--slow model, and the high-quality UVES observations appear as solid black lines.}
\label{fig:fit_HD79186}
\end{figure}

\begin{figure}[ht]
\centering
   \includegraphics[width=\hsize]{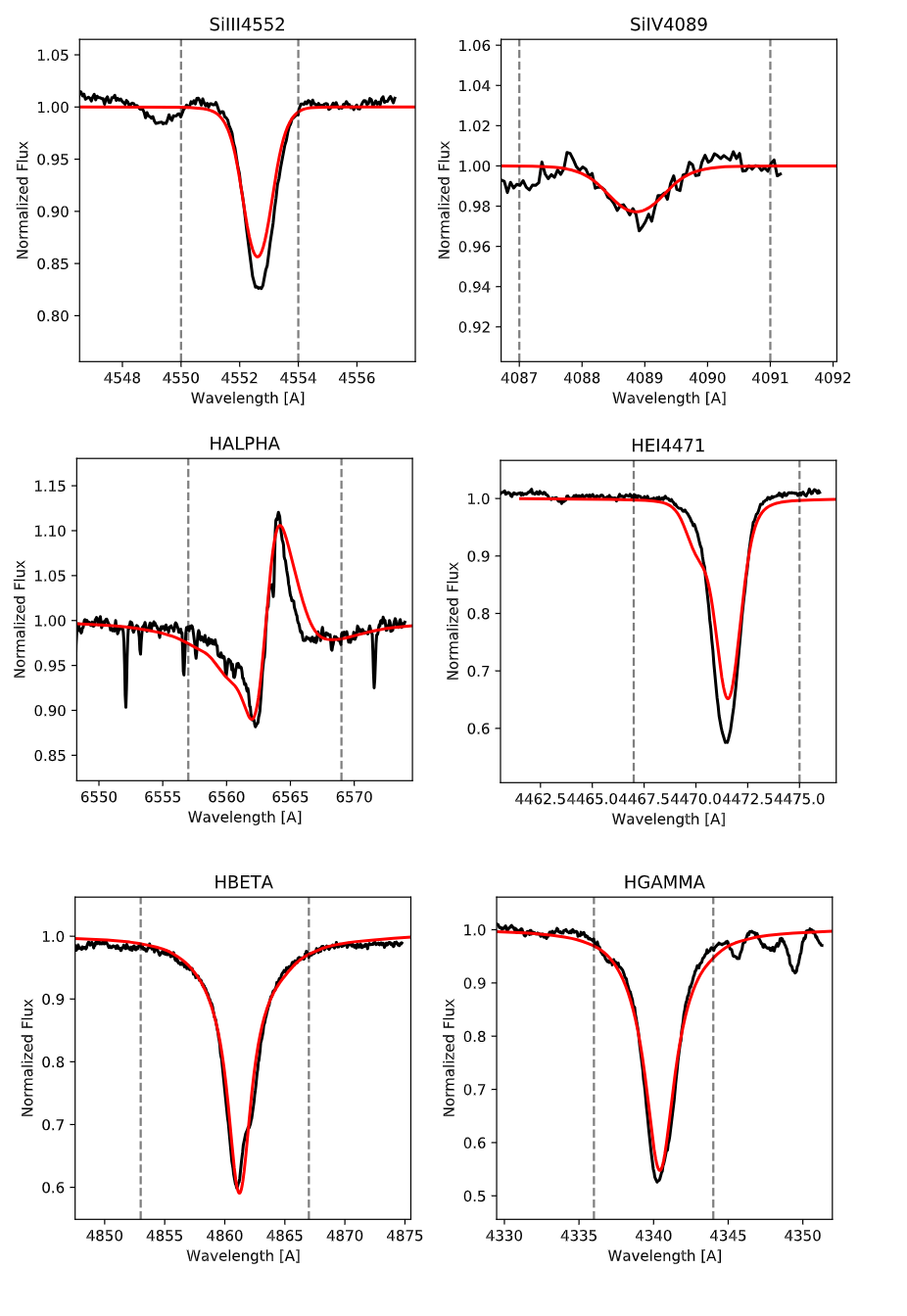}
\caption{Same as Fig.~\ref{fig:fit_HD74280}, but for HD~53138 (B3\,Ib, ESO--UVES). 
The synthetic $\delta$--slow model is shown as a solid red line, compared to the UVES observational data in black. 
Line styles follow those used in Fig.~1.}
\label{fig:fit_HD53138}
\end{figure}

In Figs.~\ref{fig:fit_HD74280} to \ref{fig:fit_HD53138}, we present representative examples of the fitting performance across different datasets and luminosity classes. Each panel illustrates the observed spectra (black) together with the best-fitting synthetic models (red) from the ISOSCELES grid, covering both photospheric and wind-sensitive diagnostic lines. Dashed vertical lines delineate the region where the $\chi^2$ was calculated. The fits generally reproduce the line cores and wings well, including features such as P-Cygni profiles, extended H$\alpha$ emission components, and helium asymmetries. Minor residuals in some cases are likely related to local continuum uncertainties or incomplete sampling of specific $\delta$ values in the grid.  

\object{HD 74280} (Fig.~\ref{fig:fit_HD74280}) is a B4\,V star whose photospheric lines (H$\beta$, H$\gamma$, and He\,\textsc{i}) are slightly broadened by rotation and well defined, with an accurately normalised continuum. The Si\,\textsc{iii}\,$\lambda4552$ line is very weak in the observed spectrum and therefore provides only limited constraints on the silicon abundance. The best-fitting fast model reproduces the observed optical features consistently, requiring a low mass-loss rate of $\dot{M} = 0.0015\times10^{-6}\,M_\odot\,\text{yr}^{-1}$. The terminal velocity inferred from the model, $v_\infty \simeq 2000$\,km/s, should be regarded as an upper-limit estimate, since the optical diagnostic lines in dwarf stars remain in absorption and are only weakly sensitive to the outer wind velocity structure.
It is important to note that wind parameters derived for dwarf stars are intrinsically more uncertain, since their diagnostic lines remain in absorption and are therefore less sensitive to wind density and velocity structure.

Fig.~\ref{fig:fit_HD206165} shows \object{HD~206165} (B2~Ib), for which the $\chi^{2}$-based fitting procedure selects a $\delta$--slow solution as the best model capable of simultaneously reproducing the extended H$\alpha$ wings. This model yields $T_{\rm eff}=20\,500$~K and $\log g=2.70$, with $\dot{M}=1.0\times10^{-6}$~M$_\odot$~yr$^{-1}$ and $v_\infty=267$~km/s. The He\,\textsc{i}$\,\lambda6678$ and Si\,\textsc{iii}$\,\lambda4552$ lines are also well reproduced, with only minor discrepancies near the line cores, possibly related to mild wind variability.

The derived terminal velocity is in very good agreement with the value reported by \citet{Markova2008}, who obtained $v_\infty\sim260$~km/s using empirical $\beta$--law parametrisations. However, our derived mass-loss rate is significantly higher than their estimate of $\dot{M}\sim(0.1$ - $0.3) \times 10^{-6}$~M$_\odot$~yr$^{-1}$. These differences are not unexpected, as their analysis relied on a different set of spectral diagnostics and adopted a prescribed velocity law, whereas our approach is based on hydrodynamically consistent wind solutions that reproduce the extended H$\alpha$ wings.

In Fig.~\ref{fig:fit_HD99953}, the object \object{HD~99953} (B1~Ib) is presented. This star was modelled using a $\delta$--slow solution, which provides a fairly good reproduction of the Balmer and helium lines. The H$\alpha$ profile shows a strong emission component with a moderately deep absorption trough, matched by a model with $\delta=0.34$, $\dot{M}=0.0398\times10^{-6}$~M$_\odot$~yr$^{-1}$, and $v_\infty=254$~km~s$^{-1}$. The line cores in Si\,\textsc{iii}$\,\lambda4552$ and He\,\textsc{i}$\,\lambda4471$ are slightly narrower in the model.

In Fig.~\ref{fig:fit_HD75149}, \object{HD 75149} (B3\,Ia) exhibits an H$\alpha$ absorption profile modelled with a P-Cygni profile, with a deep blue-shifted absorption and moderate emission component. The $\delta$--slow solution with $\dot{M} = 0.0199 \times10^{-6}\,M_\odot\,\text{yr}^{-1}$ and $v_\infty = 228$\,km\,s$^{-1}$ reproduces the shape and width of the absorption trough remarkably well. The continuum in the CASLEO spectrum is noisier in the blue range, but the Si\,\textsc{iii} and He\,\textsc{i} lines are still well matched. The resulting parameters are consistent with those of other intermediate-luminosity B supergiants in our sample.

\object{HD 79186} (Fig.~\ref{fig:fit_HD79186}), a B3\,Ib supergiant observed with ESO–UVES, displays broad H$\alpha$ wings and a weak P-Cygni profile. The best-fitting $\delta$--slow model reproduces these features with $\dot{M} = 0.0244 \times10^{-6}\,M_\odot\,\text{yr}^{-1}$ and $v_\infty = 254$\,km/s. The excellent S/N of the UVES data allows for precise continuum placement and ensures that the wind-sensitive diagnostics are robustly constrained. Within the context of our modelling, the observed profiles are best reproduced by a wind characterised by a low terminal velocity and a shallow velocity gradient. 

The H$\alpha$ line of \object{HD 53138} (B3\,Ib, Fig.~\ref{fig:fit_HD53138}) exhibits a well-developed P-Cygni profile, with extended blue absorption and strong central emission. The best-fitting model yields $\delta = 0.34$, $\dot{M} = 0.0199\times10^{-6}\,M_\odot\,\text{yr}^{-1}$, and $v_\infty = 218$\,km\,s$^{-1}$. The He\,\textsc{i} and Si\,\textsc{iii} lines are fitted within the observational uncertainties, with only minor differences in the core. The good agreement between the model and the observations supports the reliability of the $\delta$--slow solution for mid-B supergiants and reinforces the overall consistency of our fitting procedure across datasets of varying quality and resolution.

\begin{table*}
\centering
\caption{Stellar and wind parameters of the representative stars shown in Figs.~1--6.
Stellar parameters are listed first, followed by the wind parameters, adopting the same
ordering and definitions as in Tables~\ref{tab:stellar_params} and~\ref{tab:wind_params}.}
\label{tab:example_stars}
\begin{tabular}{lcccccccccccc}
\hline\hline
Star & SpT & $T_{\rm eff}$ & $\log g$ & $\log \epsilon_{\rm Si}$ &
$\alpha$ & $k$ & $\delta$ &
$\dot{M}$ & $v_\infty$ &
$v\sin i$ & $v_{\rm mac}$ & $v_{\rm esc}$ \\
 &  & [K] & [dex] &  &
 &  &  &
[$10^{-6}\,M_\odot\,{\rm yr}^{-1}$] & [km/s] &
[km/s] & [km/s] & [km/s] \\
\hline
HD~74280  & B4\,V  & 16500 & 4.05 & 7.51 & 0.61 & 0.45 & 0.10 & 0.0015 & 2031 & 76 & 19 & 855 \\
HD~206165 & B2\,Ib & 20500 & 2.70 & 7.21 & 0.45 & 0.30 & 0.30 & 1.085  & 267  & 62 & 36 & 411 \\
HD~99953  & B2\,Ia & 18500 & 2.55 & 7.51 & 0.53 & 0.15 & 0.34 & 0.0398 & 278  & 53 & 61 & 378 \\
HD~75149  & B2\,II & 16500 & 2.25 & 7.51 & 0.51 & 0.20 & 0.33 & 0.0199 & 228  & 49 & 52 & 314 \\
HD~79186  & B3\,Ib & 18500 & 2.40 & 7.51 & 0.53 & 0.15 & 0.34 & 0.0244 & 254  & 55 & 12 & 340 \\
HD~53138  & B3\,Ib & 18500 & 2.40 & 7.51 & 0.45 & 0.25 & 0.32 & 0.0199 & 218  & 63 & 27 & 339 \\
\hline
\end{tabular}
\end{table*}

Table~\ref{tab:example_stars} summarises the stellar and wind parameters of the representative B-type stars discussed in detail throughout the paper and shown in the example figures. The listed values correspond to the specific observational datasets used in each figure and provide a concise overview of the parameter range covered by the illustrative cases.

The resulting photospheric and wind parameters for the complete sample, derived with the fitting pipeline described in Sect.~\ref{sec:method}, are presented in Tables~\ref{tab:stellar_params} and~\ref{tab:wind_params}. For clarity, both tables are organised according to luminosity class, grouping stars into supergiants (I), giants and bright giants (II--III), and subgiants and dwarfs (IV--V). Table~\ref{tab:stellar_params} summarises the stellar parameters, including spectral type, effective temperature, surface gravity, silicon abundance, and the observation date and instrument. Table~\ref{tab:wind_params} lists the wind parameters, namely the mass-loss rate, terminal velocity, and the line-force parameters ($\alpha$, $k$, and $\delta$) obtained from the hydrodynamic solutions, together with the adopted microturbulent, macroturbulent, projected rotational, and escape velocities.

For stars with multiple observations obtained from different instruments (e.g. REOSC and UVES), we note that the derived photospheric and wind parameters show small but non-negligible differences. This behaviour is expected for several reasons. First, many B-type supergiants are known to exhibit line-profile variability, and the spectra analysed here correspond to different observational epochs, which can lead to variations in the inferred parameters. Second, some objects have been assigned slightly different spectral classifications in the literature, reflecting intrinsic variability and classification uncertainties, which may also contribute to differences in the best-fitting models.

In addition, the discrete sampling of the ISOSCELES grid and the intrinsic degeneracies of optical diagnostics can introduce small variations in the selected parameters even when applying the same fitting procedure. Differences in instrumental resolution are accounted for by convolving the synthetic spectra to match the resolving power of each observation. Furthermore, observational uncertainties are incorporated via the $\chi^2$ metric, in which each spectral line is weighted according to the characteristic S/N of the observations. We have verified that, within the range of data quality in our sample, no systematic trends are observed as a function of spectral resolution or S/N. Therefore, strictly homogeneous input data are not required, although minor differences between datasets are expected when comparing individual objects observed at different epochs.

\subsection{Trends in stellar and wind parameters}

To investigate systematic behaviour in our sample, we first examined how the line-force parameter $\delta$ and the global wind properties vary with effective temperature. We focus specifically on $\delta$ because this parameter regulates the sensitivity of the radiative acceleration to changes in the ionisation state of the wind, and it is the key physical quantity that distinguishes the classical fast solution from the $\delta$--slow hydrodynamical regime. Previous theoretical work has shown that increasing $\delta$ leads to slower and denser winds \citep{Cure2011}. Our fitting results indicate that supergiants systematically require higher values of $\delta$ to reproduce their observed optical spectral morphologies. It is important to stress, however, that optical absorption lines are not directly sensitive to the terminal wind velocity $v_\infty$. In our approach, $v_\infty$ is not independently measured from the optical spectra, but instead emerges consistently from the hydrodynamical solution that best reproduces the overall line profiles, including their width, depth, and wind–photosphere coupling. This behaviour produces a clear dichotomy in our sample: dwarfs are preferentially described by the classical fast solution, while giants and supergiants are predominantly reproduced by $\delta$--slow models characterised by lower terminal velocities and higher wind densities. Discrepancies between terminal velocities derived from optical modelling and higher values inferred from UV diagnostics are well known in the literature and are not specific to the present hydrodynamic framework, but also occur in empirical $\ beta$-law analyses. These differences likely reflect the fact that UV resonance lines probe the outermost wind regions and are sensitive to additional physical effects not constrained by optical diagnostics alone.

Uncertainties on the derived parameters are estimated following the approach described in Paper~I (Appendix~B), based on the distribution of the best-fitting models within the ISOSCELES grid. In this framework, uncertainties reflect the dispersion of models with similar $\chi^2$ values and therefore indicate the local density and degeneracy of the grid rather than formal statistical errors. Typical uncertainties are of the order of $\sim$1000--1500 K in $T_{\rm eff}$, $\sim$0.2--0.3 dex in $\log g$, and can be significantly larger for wind parameters such as $\dot{M}$ and $v_{\infty}$ due to their stronger degeneracy (see Paper~I for details).

Thus, analysing the behaviour of $\delta$ across the HR diagram provides a useful framework for exploring the likely wind regime and interpreting systematic differences between dwarfs and supergiants within the adopted modelling approach, rather than offering a definitive classification in individual cases. 
In all the following plots, luminosity classes are shown with different symbols and colours to highlight luminosity-class dependencies.

\begin{figure}[ht]
\centering
\includegraphics[width=1\linewidth]{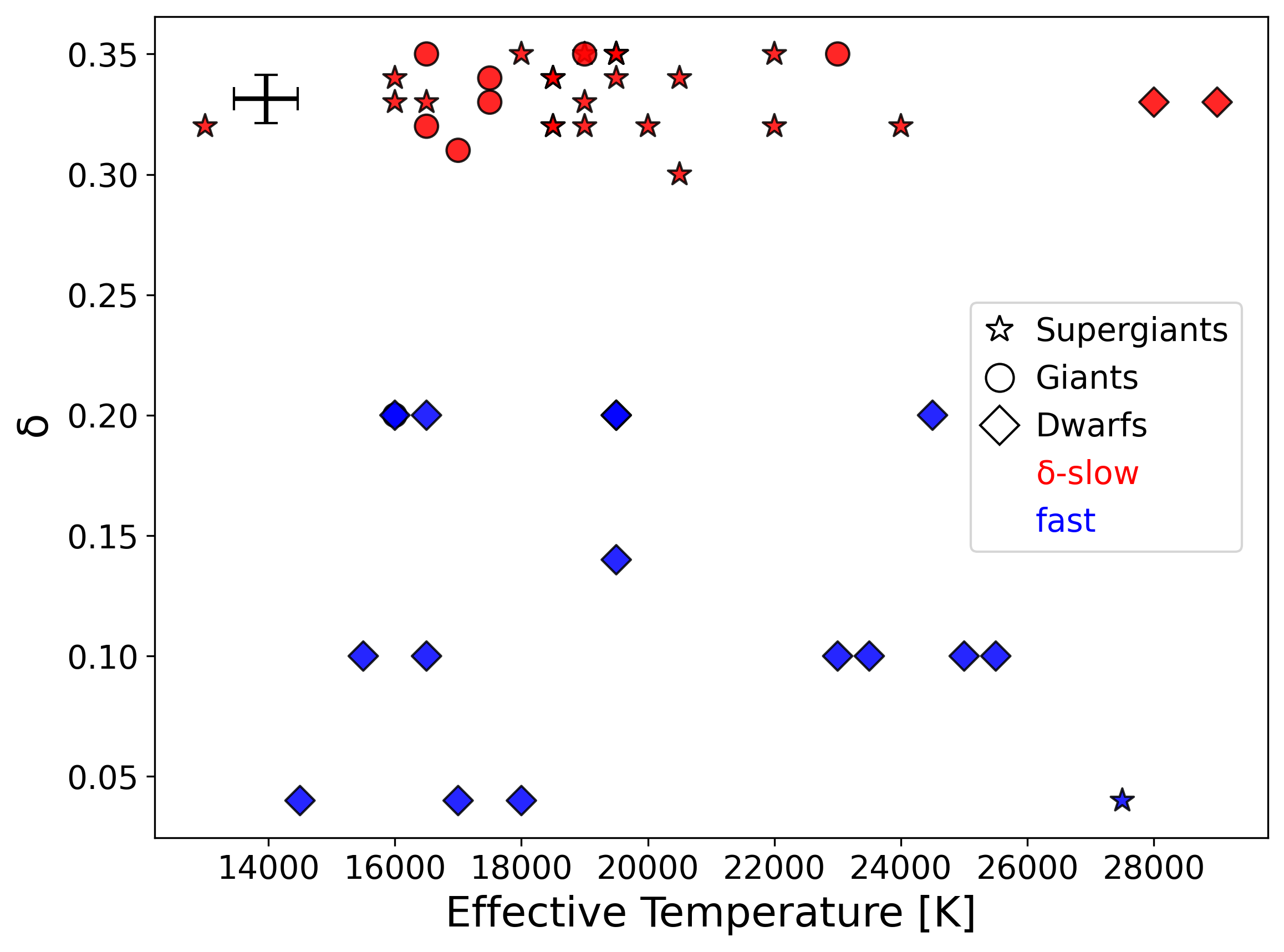}
\caption{Line-force parameter $\delta$ vs.\ effective temperature $T_{\rm eff}$. 
Symbols correspond to individual stars; colour encodes the hydrodynamical solution (red is for $\delta$--slow and blue for fast solution). 
Marker shape encodes the luminosity class group: triangles for Ia, Ib, Iab, or Ia+, circles for classes II and III, and diamonds for classes IV and V. 
All plotted quantities and sample selection are described in Sect.~\ref{sec:method} and Tables~\ref{tab:stellar_params} and ~\ref{tab:wind_params}.
Representative uncertainties are $\pm 500$ K in $T_{\rm eff}$ and $\pm 0.01$ in $\delta$ (see Paper I for details).}
\label{fig:delta_teff}
\end{figure}

Figure~\ref{fig:delta_teff} shows the distribution of the ionisation parameter $\delta$ as a function of $T_{\rm eff}$. The plot indicates that, within our fitting framework, B-type supergiants are typically characterised by higher $\delta$ values than dwarfs at comparable effective temperatures. This trend reflects the outcome of the spectral fitting procedure, in which larger $\delta$ values are preferentially selected for evolved stars in order to reproduce their observed optical line profiles, in particular the extended H$\alpha$ wings. Higher $\delta$ values correspond to a stronger ionisation dependence of the line force and are associated, in the context of the adopted hydrodynamical solutions, with slower wind acceleration and denser outflows. We stress that Fig.~\ref{fig:delta_teff} does not constitute an independent validation of the $\delta$--slow regime, but rather illustrates the systematic behaviour of the best-fitting parameters across the sample.

\begin{figure}[ht]
\centering
\includegraphics[width=0.95\linewidth]{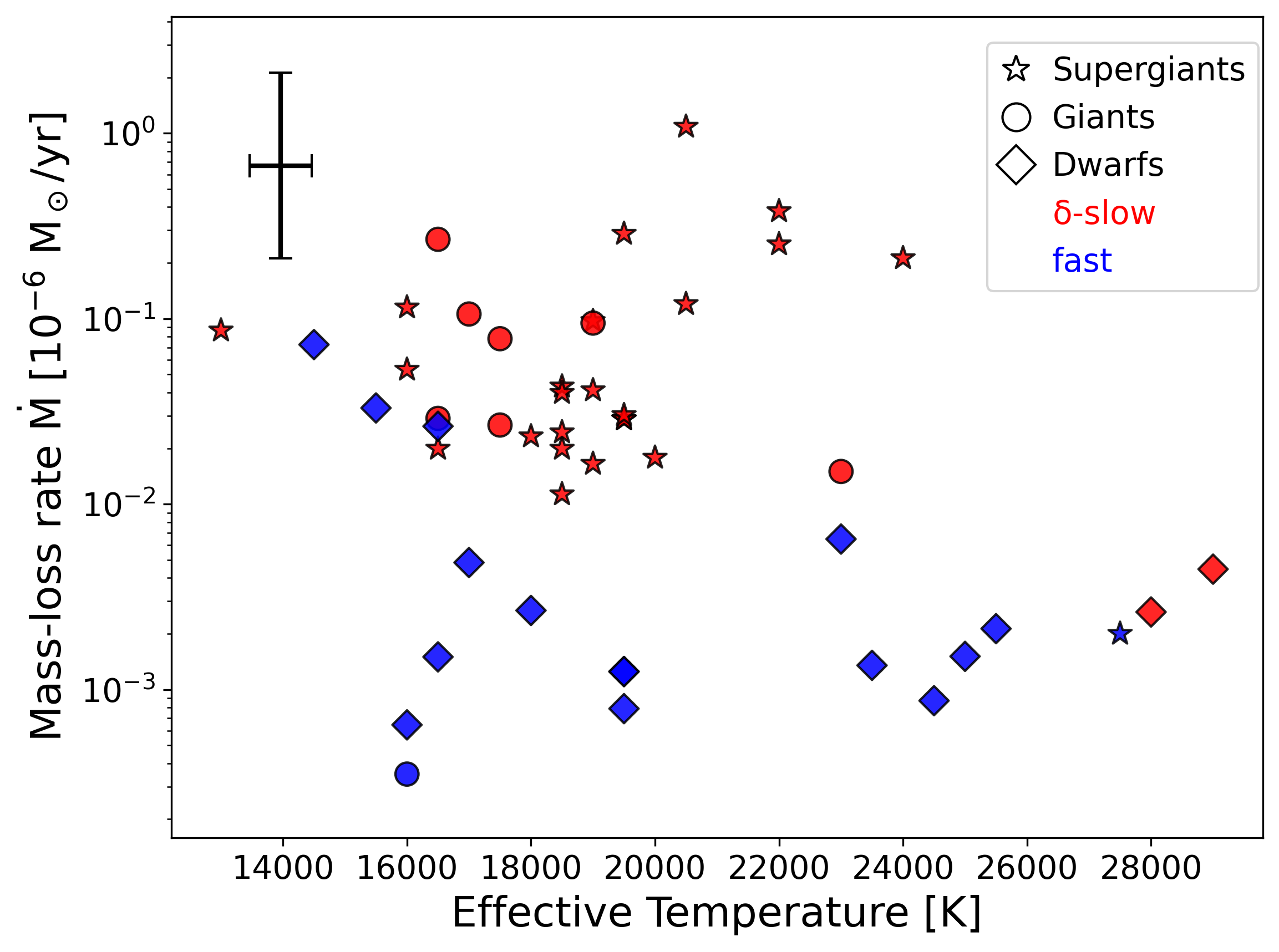}
\caption{Mass-loss rate $\dot{M}$ (units $10^{-6}\,M_\odot\,{\rm yr^{-1}}$) as a function of effective temperature $T_{\rm eff}$. 
Colours and marker shapes follow the same conventions as in Fig.~7.
Representative uncertainties are $\pm 500$ K in $T_{\rm eff}$ and $\pm 0.5$ dex in $\log \dot{M}$ (see Paper I for details).}
\label{fig:mdot_teff}
\end{figure}

\begin{figure}[ht]
\centering
\includegraphics[width=0.95\linewidth]{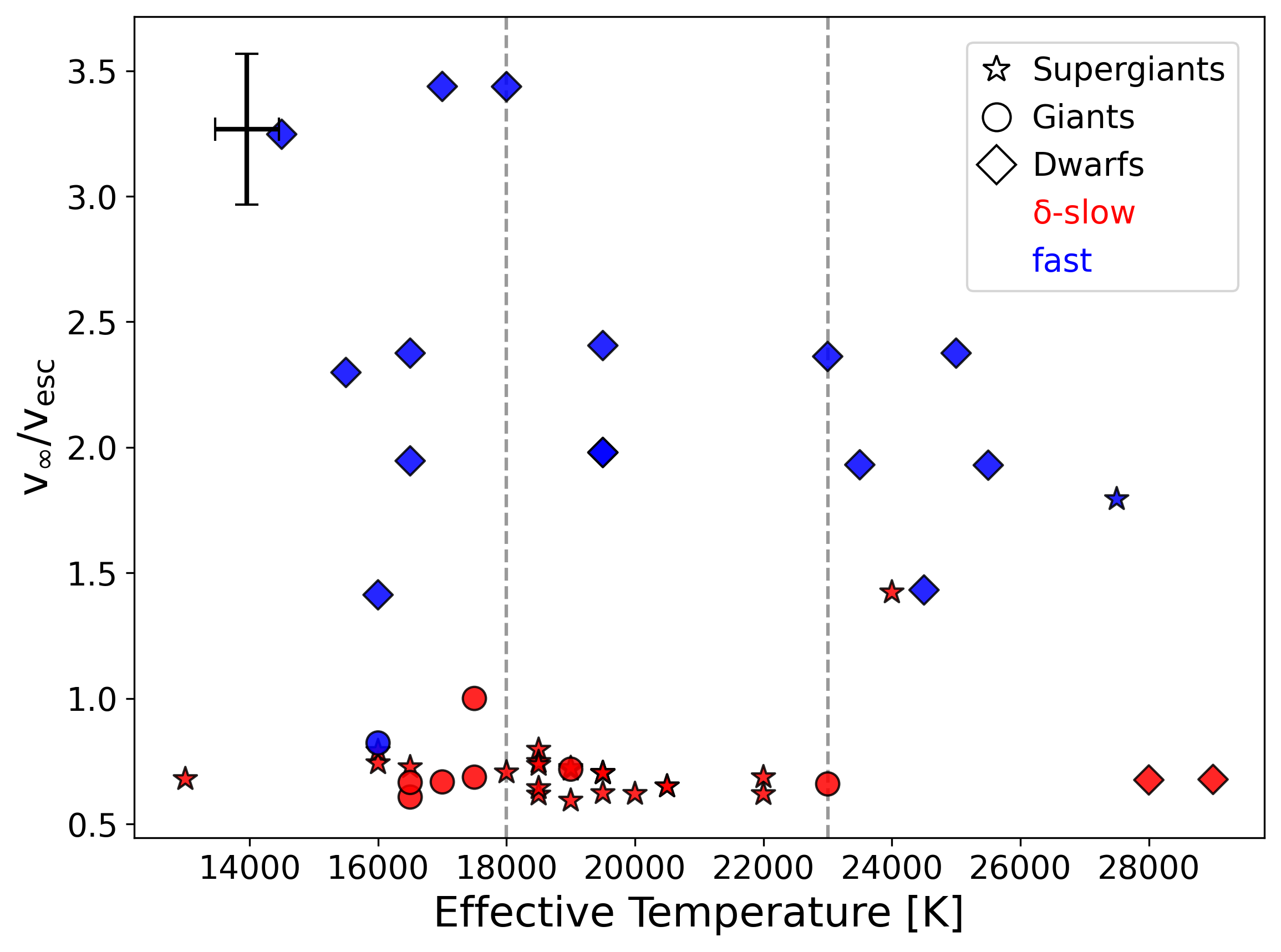}
\caption{Ratio $v_\infty/v_{\rm esc}$ as a function of effective temperature $T_{\rm eff}$. 
Colours and marker shapes follow the same conventions as in Fig.~7. 
Two vertical dashed black lines indicate the bi-stability boundaries adopted from \citet{Markova2008}. 
The clear separation into two branches reflecting the distinction between the fast and $\delta$--slow hydrodynamical regimes is discussed in Sect.~\ref{sec:results}.
Representative uncertainties are $\pm 500$ K in $T_{\rm eff}$ and $\pm 0.3$ in $v_{\infty}/v_{\rm esc}$ (see Paper I for details).}
\label{fig:vinf_teff}
\end{figure}

We also analysed how the main wind properties vary with $T_{\rm eff}$. Figures~\ref{fig:mdot_teff} and~\ref{fig:vinf_teff} present the distributions of mass-loss rates and terminal velocities, respectively. As before, dwarfs and supergiants are distinguished by different colours and symbols.

As expected, Fig.~\ref{fig:mdot_teff} shows that supergiants exhibit systematically higher mass-loss rates than dwarfs at similar effective temperatures. This behaviour reflects their lower surface gravities and larger radii, which modify the atmospheric hydrodynamic structure and favour the development of denser winds. In addition to this global separation between luminosity classes, the distribution of $\dot{M}$ shows a noticeable change around the temperature range associated with the classical bi-stability region, suggesting the presence of a discontinuity in the mass-loss rates. 
It is worth noting that the apparent discontinuity in $\dot{M}$ is driven primarily by the supergiant population, whereas dwarfs and subgiants display a more continuous behaviour across the same temperature range. This is consistent with previous studies that reported a bi-stability-related change in mass-loss rates for evolved B-type stars \citep[e.g.,][]{Haucke2018}, whereas similar signatures are less evident or absent in less-evolved objects. In contrast, the behaviour of $v_\infty$ inferred from the present optical analysis does not show a corresponding abrupt transition, and therefore differs from UV-based determinations reported in earlier works. This contrast highlights that different wind diagnostics are sensitive to different regions of the outflow and may respond differently to changes in the line-driving conditions.

Figure~\ref{fig:vinf_teff} shows the behaviour of the wind through the ratio $v_{\infty}/v_{\rm esc}$ as a function of effective temperature. The escape velocity was computed as 
\[
v_{\rm esc} = \sqrt{\frac{2GM(1-\Gamma)}{R_*}},
\]
where $\Gamma$ is the Eddington factor due to electron (Thomson) scattering, defined as the ratio of radiative to gravitational acceleration, thus accounting for the reduction of the effective gravity by radiation pressure. Two distinct groups are apparent: stars converging to the classical fast solution populate an upper branch with larger $v_{\infty}/v_{\rm esc}$ values, while objects described by $\delta$--slow solutions occupy a lower branch, typically with $v_{\infty}/v_{\rm esc} \lesssim 1.5$. In our framework, this separation reflects the underlying hydrodynamical regime rather than a monotonic dependence on effective temperature alone.

The two vertical dashed lines indicate the bi-stability boundaries reported by \citet[][their Fig.~12]{Markova2008}, which were identified from analyses based primarily on UV-derived terminal velocities and empirical $\beta$--law modelling. While such approaches commonly recover a sharp drop in $v_{\infty}$ around $T_{\rm eff}\sim 21\,000$~K, our results do not show an abrupt temperature-driven jump. Instead, the observed structure in $v_{\infty}/v_{\rm esc}$ arises from the coexistence of fast and $\delta$--slow hydrodynamical solutions over a similar temperature range.

This suggests that the classical bistability region overlaps with the parameter space in which multiple wind regimes can coexist. A detailed, homogeneous comparison between optical and UV diagnostics is required to fully assess the origin of the bi-stability jump, and is beyond the scope of the present work.

\begin{figure}[ht]
\centering
\includegraphics[width=0.95\linewidth]{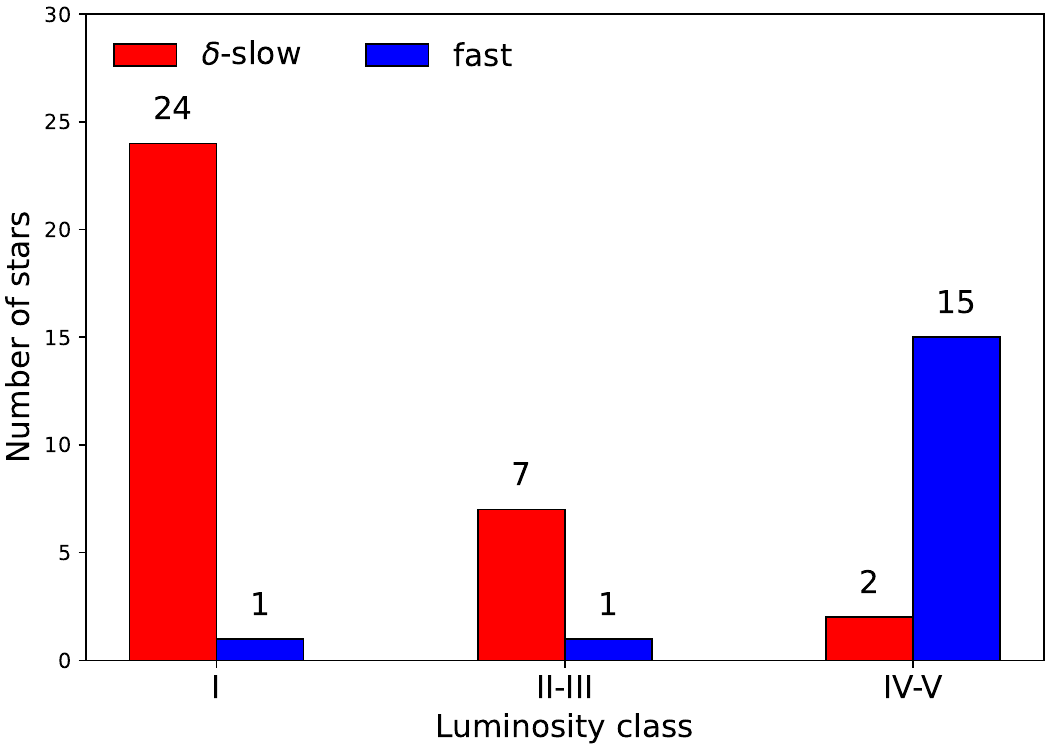}
\caption{Distribution of the 50 B-type spectra analysed in this work according to luminosity class and hydrodynamical solution. Blue bars correspond to fast solutions, red bars to $\delta$--slow solutions.}
\label{fig:bar}
\end{figure}

Figure~\ref{fig:bar} summarises the global distribution of wind regimes across luminosity classes. The histogram shows a clear dichotomy: nearly all B supergiants and giants (classes~I–III) are best reproduced by $\delta$--slow solutions, while most subgiants and dwarfs (classes~IV–V) correspond to the classical fast solution. Quantitatively, 94\% of the evolved stars in our sample fall in the $\delta$--slow category, whereas only two of the dwarfs do. In contrast to analyses based on prescribed $\beta$ velocity laws, the hydrodynamically consistent velocity fields adopted here may reduce the ambiguities often encountered in standard modelling. In particular, empirical $\beta$-law fits can reproduce similar H$\alpha$ morphologies with different velocity structures, sometimes predicting pure emission profiles with P-Cygni features, and sometimes predicting pure absorption profiles with P-Cygni features. By deriving the velocity field self-consistently from the hydrodynamical solution, the models provide a more physically constrained interpretation of the observed line profiles.

This segregation suggests that the $\delta$ parameter governs the transition between the two hydrodynamical regimes. As stars evolve off the main sequence and their surface gravities decrease, models with larger $\delta$ values (reflecting changes in ionisation throughout the wind) reproduce the observed wind diagnostics more accurately, yielding slower, denser outflows. The separation seen in Fig.~\ref{fig:bar} is therefore consistent with theoretical predictions of the $\delta$--slow solution and supports its relevance as a viable wind regime for B-type supergiants.

\section{Discussion and conclusions}\label{sec:dis}

\subsection{Physical implications of the $\delta$--slow solution}

Our analysis suggests a systematic trend, within the adopted modelling framework, between the inferred wind regime and the evolutionary state of B-type stars across the analysed sample. In particular, most evolved objects in our sample tend to be better reproduced by $\delta$--slow solutions, whereas the majority of dwarfs and subgiants are preferentially described by classical fast solutions. The distinction between the two regimes is based on the global quality of fits to the optical diagnostics, primarily the H$\alpha$ profile and its coupling with photospheric lines, as quantified by $\chi^{2}$ minimisation. We note, however, that optical diagnostics such as H$\alpha$ have limited sensitivity to the outer wind regions and therefore provide only indirect constraints on the terminal velocity. In some cases, this may lead to discrepancies with terminal velocities inferred from ultraviolet diagnostics, which probe the outer wind layers.

It is important to emphasise, however, that for individual stars both fast and $\delta$--slow solutions can in some cases provide comparably good representations of the observed optical profiles, as already discussed by \citet{venero2016}. Our results, therefore, do not eliminate this intrinsic degeneracy at the level of single objects. Instead, they indicate that, when considered statistically over a relatively large and homogeneous sample, evolved B-type stars tend to favour $\delta$--slow solutions more frequently, while hotter and less evolved objects are more often consistent with fast solutions. This behaviour is qualitatively consistent with the theoretical expectations of \citet{Cure2011}, although a strict one-to-one correspondence between evolutionary stage and wind regime cannot be established from optical diagnostics alone.

Within the m-CAK framework, the $\delta$--slow solution arises when the ionisation parameter $\delta$ is increased, modifying the sensitivity of the radiative acceleration to the local ionisation balance. This results in a more gradual wind acceleration and higher densities near the wind base. Within our modelling framework, such wind structures tend to reproduce the observed optical spectral morphology of B-type giants and supergiants more accurately, particularly the extended H$\alpha$ wings. In contrast, dwarf stars, characterised by higher surface gravities and more compact atmospheres, are more frequently reproduced by models that converge to the classical fast solution, which, within the adopted hydrodynamic models, is associated with steeper acceleration and lower wind densities.
These associations should be interpreted as statistical trends derived from optical diagnostics, rather than as unique or definitive physical solutions, since different wind structures can in some cases produce similar optical line profiles. In particular, the wind structure inferred from optical diagnostics may predominantly reflect the conditions in the inner wind and may not fully capture the behaviour of the outermost regions.

\subsection{Reliability of hydrodynamical models}

The fitting procedure based on the ISOSCELES grid yields internally consistent parameter estimates across the analysed sample. By employing hydrodynamical wind structures rather than prescribing empirical velocity laws, the mass-loss rate, terminal velocity, and velocity gradient are derived self-consistently within the adopted m-CAK framework for a given set of line-force parameters ($\alpha$, $k$, and $\delta$). For a given star, solutions associated with the wind regime that best reproduces the optical diagnostics (fast or $\delta$-slow) tend to yield lower $\chi^{2}$ values than alternative solutions, indicating better internal consistency of the fits within this modelling approach.

\begin{figure}
\centering
\includegraphics[width=0.99\linewidth]{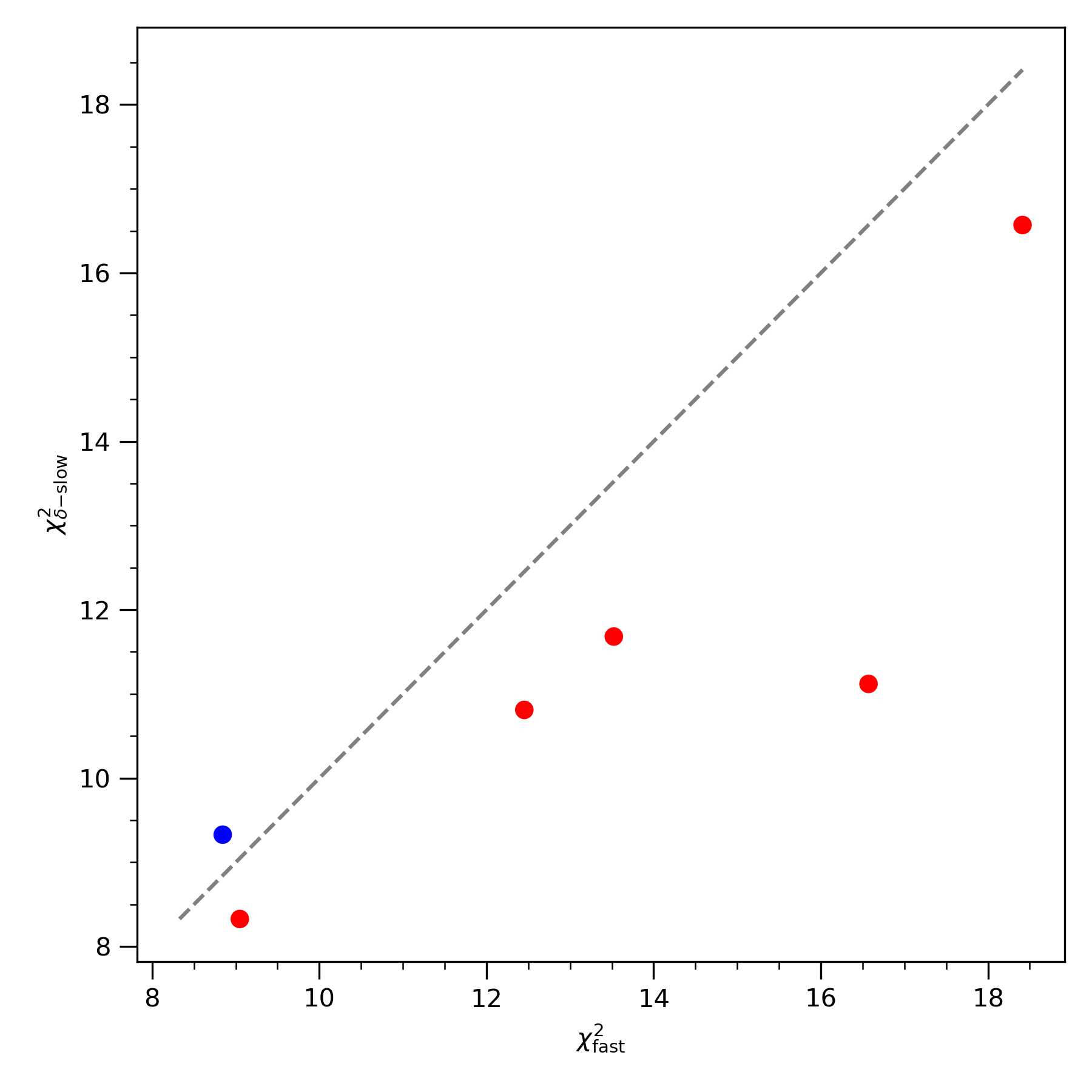}
\caption{Comparison between the minimum $\chi^2$ values obtained from fast and $\delta$--slow hydrodynamical solutions for the stars shown as spectral fitting examples in Sect.~\ref{sec:results}. For each star, the lowest $\chi^2$ obtained among fast models and among $\delta$--slow models is displayed. The dashed line indicates the one-to-one relation. Points below this line correspond to cases where $\delta$--slow solutions provide better fits, while points above the line favour fast solutions.}
\label{fig:chi2_fast_delta}
\end{figure}

To illustrate this behaviour for the representative stars discussed in Sect.~\ref{sec:results}, Fig.~\ref{fig:chi2_fast_delta} compares the minimum $\chi^2$ values obtained within the fast and $\delta$--slow solution families. For each star, the lowest $\chi^2$ obtained among models converging to the classical fast solution and among models corresponding to $\delta$--slow solutions was identified within the explored ISOSCELES grid. Each point, therefore, represents the best-fitting model within each wind regime. Points located below the one-to-one relation correspond to stars for which the $\delta$--slow solution provides the lower $\chi^2$, while points above the line indicate cases where the fast solution provides the better fit. For the majority of the examples shown here, the minimum $\chi^2$ is obtained for $\delta$--slow models, consistent with the spectral morphologies discussed in Sect.~\ref{sec:results}. This comparison therefore supports the internal consistency of the adopted wind regimes for the illustrative cases analysed in this work.

These results highlight the differences between empirical $\beta$--law parametrisations and hydrodynamically motivated wind solutions in modelling the optical spectra of evolved B-type stars. In several previous studies, large $\beta$ values have been adopted in empirical $\beta$--laws to reproduce the broad H$\alpha$ wings observed in B supergiants. While such prescriptions can successfully match the observed profiles, they rely on imposing a slowly accelerating velocity field through the chosen parametrisation rather than deriving it from the hydrodynamical equations. High-$\beta$ empirical fits have been reported in the literature (e.g., \citealt{Markova2008,petrov2014}), reflecting modelling strategies aimed at reproducing similar spectral morphologies.

In contrast, within our modelling framework, a slow acceleration emerges naturally as a $\delta$--slow hydrodynamical solution for a given set of line-force parameters ($\alpha$, $k$, and $\delta$), without explicitly tuning the velocity field. The observed optical profiles are therefore reproduced through the response of the line force to the ionisation structure of the wind, rather than by prescribing a specific velocity gradient. This provides a physically motivated interpretation for why empirical models with large $\beta$ values have been successful in reproducing similar spectral features.

This interpretation nevertheless relies on the assumption that sufficiently large $\delta$ values can be attained in evolved B-type stars. The physical origin and range of $\delta$ remain model-dependent, and independent constraints on the ionisation structure of the wind, particularly from ultraviolet diagnostics, are required to further assess the plausibility of this regime.

It should be noted that the scatter observed within each luminosity class suggests that additional effects related to stellar evolution, such as variations in surface CNO abundances or differences in wind structure, may also contribute to the observed spread in $\dot{M}$ and inferred wind properties. A systematic exploration of these effects would require dedicated model grids with varying chemical compositions and is therefore beyond the scope of the present work.

\subsection{Broader context and future directions}

The distinction between fast and $\delta$--slow winds has broader implications for evolutionary models of massive stars, feedback processes, and population synthesis studies. By accounting for systematic differences in wind properties across the B-type sequence, improved constraints can be placed on mass-loss prescriptions used in stellar evolution and feedback calculations.

Future work will extend this analysis to O-type stars and explore the impact of metallicity. In addition, combining optical diagnostics with ultraviolet spectroscopy will be essential to further constrain the ionisation balance and terminal wind properties. Infrared observations with CRIRES at the VLT, targeting the Brackett-$\alpha$ line, will also be incorporated to provide an independent diagnostic sensitive to wind density and velocity structure and to test the consistency of the inferred hydrodynamical solutions.

\subsection{Conclusions}

This work demonstrates the potential of hydrodynamically consistent modelling to reproduce the optical spectral features of B-type stars. Our main conclusions are as follows:

\begin{itemize}
\item A significant fraction of B-type supergiants in our sample are better reproduced, within the adopted modelling framework, by $\delta$--slow solutions, suggesting that slowly accelerating winds may provide an effective description of their optical spectra.
\item B-type dwarfs and subgiants are more frequently consistent with classical fast solutions, in line with their higher gravities and more compact atmospheres.
\item The use of the ISOSCELES grid provides a physically motivated alternative to empirical $\beta$--law prescriptions by deriving the wind structure from hydrodynamical solutions rather than prescribing the velocity field.
\item The observed correlation between luminosity class and preferred wind regime highlights the importance of tailored wind modelling when interpreting optical spectra of massive stars.
\end{itemize}

These results highlight the importance of incorporating hydrodynamical consistency in spectroscopic analyses, while also emphasising the need for multiwavelength diagnostics to fully constrain the physical properties of stellar winds. However, given the limitations of optical diagnostics in constraining the outer wind regions, these conclusions should be regarded as indicative, and further constraints from ultraviolet spectroscopy are required for a definitive assessment of the wind regime. In particular, multi-wavelength analyses will be necessary to assess whether a single hydrodynamical solution can consistently describe both the inner and outer wind regions.

\begin{acknowledgements}
The authors thank J. Puls for adapting the \textsc{Fastwind} code to include the hydrodynamical solution from \textsc{Hydwind}. N.M. thanks the support from ANID BECAS/DOCTORADO NACIONAL 21221364. I.A., C.A. and M.C. are grateful for the support from the Fondecyt projects 1230131 and 1261498. MC \& CA acknowledge the support from Centro de Astrof\'isica de Valpara\'iso.
ROJV acknowledges financial support from CONICET (PIP 11220200101337CO) and the Universidad Nacional de La Plata (Programa de Incentivos 11/G192 and 11/G193). A.L. acknowledges in part funding by the Belgian Federal Science Policy Office - Policy for Science Program Contract No. P4S/251/Gaia-BRASS.
This project has been partially co-funded by the European Union, Project 101183150 - OCEANS.
This work has been made possible thanks to the use of AWS-U.Chile-NLHPC credits. Powered@NLHPC: This research was partially supported by the NLHPC's supercomputing infrastructure (CCSS210001).
\end{acknowledgements}

   \bibliographystyle{aa} 
   \bibliography{biblio} 

\begin{appendix}
\onecolumn

\section{Pipeline structure}

\begin{figure*}[h!]
    \centering
    \includegraphics[width=0.8\linewidth]{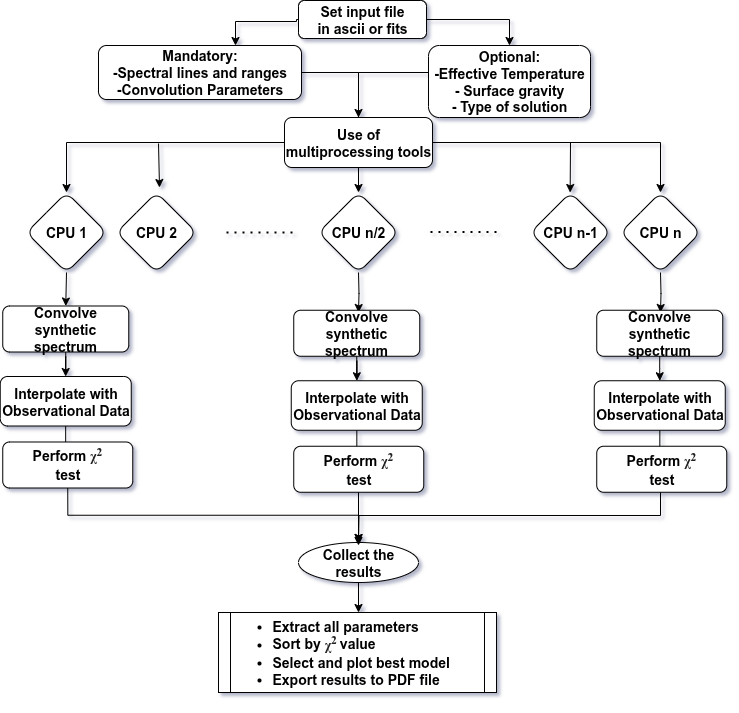}
    \caption{Flowchart of the spectral fitting parallel procedure. 
    }
    \label{fig:code_diagram}
\end{figure*}

\onecolumn
\section{Obtained parameters for all 50 stars}
\begin{table*}[h]
\centering
\caption{Stellar parameters of the studied B-type stars grouped according to luminosity class.}
\label{tab:stellar_params}
\begin{tabular}{llrrrll}
\hline
Star & Spectral Type & $T_{\rm eff}$ [K] & $\log g$ [dex] & $\log \varepsilon_{\rm Si}$ & Observation date & Instrument \\
\hline

\multicolumn{7}{c}{Luminosity class I} \\
\hline
HD2905   & BC0.7 Ia & 22000 & 3.00 & 7.51 & 2016-07-08 & HERMES \\
HD24398  & B1 Ib    & 22000 & 3.00 & 7.21 & 2009-11-12 & FIES \\
HD34085  & B8 Ia    & 13000 & 1.95 & 7.21 & 2006-01-15 & REOSC \\
HD37128  & B0 Ia    & 19500 & 2.55 & 7.51 & 2014-02-11 & HERMES \\
HD38771  & B0.5 Ia  & 24500 & 3.15 & 7.51 & 2017-12-10 & HERMES \\
HD41117  & B2 Iab   & 19000 & 2.55 & 7.81 & 2006-01-15 & REOSC \\
HD47240  & B1 Ib    & 18000 & 2.40 & 7.21 & 2006-01-15 & REOSC \\
HD53138  & B3 Ia    & 18500 & 2.70 & 7.51 & 2006-01-15 & REOSC \\
HD53138  & B3 Ia    & 18500 & 2.40 & 7.51 & 2014-09-16 & UVES \\
HD54764  & B1 Ib    & 20000 & 2.85 & 7.51 & 2006-02-08 & FEROS \\
HD75149  & B3 Ia    & 16500 & 2.25 & 7.51 & 2013-02-07 & REOSC \\
HD75149  & B5 Iab   & 19000 & 2.55 & 7.36 & 2016-09-21 & UVES \\
HD79186  & B5 Ia    & 16000 & 2.10 & 7.51 & 2006-01-15 & REOSC \\
HD79186  & B5 Ia    & 18500 & 2.40 & 7.51 & 2020-06-12 & UVES \\
HD92964  & B2.5 Ia  & 19500 & 2.55 & 7.36 & 2013-02-05 & REOSC \\
HD99953  & B2 Ia    & 18500 & 2.55 & 7.51 & 2015-02-13 & REOSC \\
HD99953  & B2 Ia    & 19500 & 2.55 & 7.21 & 2020-06-09 & UVES \\
HD109867 & B0.5 Iab & 18000 & 2.40 & 7.51 & 2021-08-27 & UVES \\
HD115842 & B0.5 Ia  & 24000 & 2.55 & 7.51 & 2014-04-11 & REOSC \\
HD115842 & B0.5 Ia  & 27500 & 2.85 & 7.81 & 2020-06-28 & UVES \\
HD152236 & B1.5 Ia+ & 20500 & 2.70 & 7.51 & 2014-09-25 & UVES \\
HD154090 & B2 Iab   & 19500 & 2.55 & 7.36 & 2020-06-15 & UVES \\
HD171012 & B0.2 Ia  & 19500 & 2.55 & 7.51 & 2021-10-04 & UVES \\
HD204172 & B0 Ib    & 19000 & 2.55 & 7.81 & 2009-11-11 & FIES \\
HD206165 & B2 Ib    & 20500 & 2.70 & 7.21 & 2009-11-10 & FIES \\
\hline
\multicolumn{7}{c}{Luminosity classes II--III} \\
\hline
HD21483  & B3 III   & 17500 & 3.45 & 7.36 & 2011-11-09 & HERMES \\
HD29248  & B2 III   & 23000 & 3.90 & 7.51 & 2013-12-17 & HERMES \\
HD31327  & B2 II    & 17000 & 2.55 & 7.51 & 2009-11-11 & FIES \\
HD186660 & B3 III   & 16500 & 3.30 & 7.51 & 2014-10-15 & FIES \\
HD198820 & B3 III   & 17500 & 3.30 & 7.36 & 2015-07-22 & HERMES \\
HD204536 & B3 III   & 16500 & 3.30 & 7.51 & 2014-10-16 & FIES \\
HD209008 & B3 III   & 16000 & 3.90 & 7.21 & 2014-10-16 & FIES \\
HD220787 & B3 III   & 19000 & 4.05 & 7.36 & 2018-08-01 & HERMES \\
\hline
\multicolumn{7}{c}{Luminosity classes IV--V} \\
\hline
HD20365  & B3 V     & 16000 & 3.75 & 7.81 & 2013-12-14 & HERMES \\
HD25204  & B3 IV    & 15500 & 3.30 & 7.81 & 2013-12-15 & HERMES \\
HD34989  & B1 V     & 25500 & 4.20 & 7.81 & 2013-02-05 & FEROS \\
HD35299  & B1.5 V   & 23500 & 4.20 & 7.21 & 2013-02-06 & FEROS \\
HD35912  & B2 V     & 19500 & 4.05 & 7.51 & 2008-11-08 & FIES \\
HD36430  & B2 V     & 19500 & 4.20 & 7.51 & 2008-11-08 & FIES \\
HD36591  & B1 IV    & 29000 & 4.20 & 7.51 & 2008-11-08 & FIES \\
HD36629  & B2 V     & 19500 & 4.05 & 7.51 & 2008-11-08 & FIES \\
HD36862  & B0.5 V   & 25000 & 4.20 & 7.51 & 2008-11-07 & FIES \\
HD36960  & B0.5 V   & 28000 & 4.20 & 7.51 & 2011-11-08 & FEROS \\
HD37481  & B1.5 V   & 22000 & 4.05 & 7.21 & 2017-10-30 & HERMES \\
HD37744  & B1.5 V   & 23000 & 4.20 & 7.51 & 2008-11-08 & FIES \\
HD41753  & B3 IV    & 16000 & 4.05 & 7.21 & 2017-04-06 & HERMES \\
HD74280  & B4 V     & 16500 & 4.05 & 7.51 & 2017-10-21 & HERMES \\
HD179406 & B3 V     & 14500 & 3.00 & 7.66 & 2013-06-25 & FIES \\
HD182568 & B3 IV    & 16500 & 3.45 & 7.51 & 2013-06-24 & FIES \\
HD218537 & B3 V     & 17000 & 4.05 & 7.51 & 2013-06-26 & FIES \\
\hline
\end{tabular}
\tablefoot{Typical uncertainties are $\pm(1.0$–$1.5)\times10^{3}$ K in $T_{\rm eff}$, $\pm0.2$ dex in $\log g$, $\pm0.2$ dex in Si abundance.}
\end{table*}

\begin{table*}
\centering
\caption{Wind parameters of the studied B-type stars grouped according to luminosity class.}
\label{tab:wind_params}
\begin{tabular}{lrrrrrrrrr}
\hline
Star & $\alpha$ & $k$ & $\delta$ & $\dot{M}$ & $v_{\infty}$ [km/s] & $v_{mic}$ [km/s] & $v\sin i$ [km/s] & $v_{mac}$ [km/s] & $v_{esc}$ [km/s] \\
\hline

\multicolumn{10}{c}{Luminosity class I} \\
\hline
HD2905   & 0.51 & 0.30 & 0.35 & 0.252   &  337 & 10 & 66  & 23 & 491 \\
HD24398  & 0.45 & 0.35 & 0.32 & 0.380   &  304 & 10 & 43  & 17 & 491 \\
HD34085  & 0.47 & 0.55 & 0.32 & 0.0864  &  184 & 10 & 31  & 56 & 271 \\
HD37128  & 0.51 & 0.20 & 0.35 & 0.287   &  263 & 10 & 48  & 11 & 374 \\
HD38771  & 0.45 & 0.35 & 0.35 & 0.0971  &  318 & 10 & 49  & 29 & 536 \\
HD41117  & 0.53 & 0.10 & 0.33 & 0.115   &  279 & 20 & 46  & 74 & 376 \\
HD47240  & 0.57 & 0.05 & 0.34 & 0.0431  &  273 & 15 & 107 & 76 & 343 \\
HD53138  & 0.45 & 0.25 & 0.32 & 0.0113  &  255 & 10 & 41  & 57 & 413 \\
HD53138  & 0.45 & 0.25 & 0.32 & 0.0199  &  218 & 10 & 63  & 27 & 339 \\
HD54764  & 0.45 & 0.35 & 0.32 & 0.0178  &  279 & 10 & 125 & 31 & 450 \\
HD75149  & 0.51 & 0.20 & 0.33 & 0.0199  &  228 & 15 & 49  & 52 & 314 \\
HD75149  & 0.51 & 0.20 & 0.33 & 0.0165  &  270 & 10 & 61  & 31 & 377 \\
HD79186  & 0.55 & 0.15 & 0.34 & 0.0531  &  223 &  5 & 42  & 52 & 282 \\
HD79186  & 0.53 & 0.15 & 0.34 & 0.0244  &  254 & 10 & 55  & 12 & 340 \\
HD92964  & 0.51 & 0.20 & 0.35 & 0.0287  &  263 &  5 & 46  & 48 & 374 \\
HD99953  & 0.53 & 0.15 & 0.34 & 0.0398  &  278 & 15 & 53  & 61 & 378 \\
HD99953  & 0.51 & 0.20 & 0.35 & 0.0287  &  263 & 10 & 84  & 33 & 375 \\
HD109867 & 0.51 & 0.20 & 0.35 & 0.0232  &  242 & 10 & 77  & 22 & 343 \\
HD115842 & 0.65 & 0.10 & 0.32 & 0.212   &  441 & 20 & 62  & 71 & 310 \\
HD115842 & 0.51 & 0.15 & 0.04 & 0.002   &  705 & 10 & 71  & 21 & 393 \\
HD152236 & 0.47 & 0.35 & 0.34 & 0.120   &  267 &  1 & 60  & 48 & 411 \\
HD154090 & 0.51 & 0.20 & 0.35 & 0.0287  &  263 & 10 & 60  & 45 & 374 \\
HD171012 & 0.45 & 0.30 & 0.34 & 0.0302  &  233 & 10 & 87  & 10 & 374 \\
HD204172 & 0.51 & 0.15 & 0.32 & 0.0411  &  272 & 10 & 58  & 16 & 376 \\
HD206165 & 0.45 & 0.30 & 0.30 & 1.085   &  267 & 10 & 62  & 36 & 411 \\

\hline
\multicolumn{10}{c}{Luminosity classes II--III} \\
\hline
HD21483  & 0.51 & 0.45 & 0.34 & 0.078   &  381 &  5 & 123 & 11 & 381 \\
HD29248  & 0.51 & 0.60 & 0.35 & 0.0150  &  488 &  1 & 34  & 40 & 740 \\
HD31327  & 0.47 & 0.40 & 0.31 & 0.106   &  253 & 10 & 35  & 25 & 379 \\
HD186660 & 0.45 & 0.60 & 0.32 & 0.268   &  283 & 10 & 15  &  8 & 466 \\
HD198820 & 0.51 & 0.60 & 0.33 & 0.0267  &  368 &  1 & 33  &  6 & 536 \\
HD204536 & 0.51 & 0.55 & 0.35 & 0.029   &  350 &  1 & 70  & 43 & 526 \\
HD209008 & 0.51 & 0.55 & 0.20 & 0.00035 &  538 &  1 & 21  & 12 & 654 \\
HD220787 & 0.55 & 0.60 & 0.35 & 0.0947  &  528 &  1 & 26  & 15 & 735 \\

\hline
\multicolumn{10}{c}{Luminosity classes IV--V} \\
\hline
HD20365  & 0.55 & 0.60 & 0.20 & 0.000645 &  873 &  1 & 125 &  5 & 618 \\
HD25204  & 0.61 & 0.55 & 0.10 & 0.0330  & 1186 & 10 & 52.5& 24 & 516 \\
HD34989  & 0.55 & 0.30 & 0.10 & 0.00213 & 1659 &  1 & 48  & 16 & 860 \\
HD35299  & 0.55 & 0.35 & 0.10 & 0.00135 & 1616 &  1 & 15  &  4 & 837 \\
HD35912  & 0.65 & 0.20 & 0.20 & 0.00125 & 1467 &  1 & 21  & 10 & 741 \\
HD36430  & 0.65 & 0.20 & 0.14 & 0.000789& 1886 &  1 & 25  & 15 & 784 \\
HD36591  & 0.51 & 0.45 & 0.33 & 0.00446 &  607 &  1 & 19  &  8 & 896 \\
HD36629  & 0.65 & 0.20 & 0.20 & 0.00125 & 1467 &  1 & 17  &  7 & 741 \\
HD36862  & 0.61 & 0.45 & 0.10 & 0.00151 & 2031 &  1 & 76  & 19 & 855 \\
HD36960  & 0.51 & 0.50 & 0.33 & 0.00262 &  599 &  5 & 35  & 25 & 887 \\
HD37481  & 0.61 & 0.35 & 0.10 & 0.00648 & 1826 &  1 & 74  & 14 & 773 \\
HD37744  & 0.55 & 0.40 & 0.20 & 0.000871& 1190 &  1 & 37  & 15 & 831 \\
HD41753  & 0.65 & 0.55 & 0.04 & 0.00267 & 2375 &  5 & 36  & 23 & 691 \\
HD74280  & 0.61 & 0.45 & 0.10 & 0.0015  & 2031 & 10 & 76  & 19 & 855 \\
HD179406 & 0.65 & 0.35 & 0.04 & 0.0725  & 1458 & 10 & 158 &  9 & 449 \\
HD182568 & 0.65 & 0.45 & 0.20 & 0.0263  & 1086 &  1 & 105 & 20 & 558 \\
HD218537 & 0.65 & 0.60 & 0.04 & 0.00484 & 2427 &  5 & 150 & 40 & 706 \\

\hline
\end{tabular}
\tablefoot{
Mass-loss rates are expressed in units of $10^{-6}$~M$_\odot$/yr. Typical uncertainties are $3$–$5$ km/s in microturbulence, $\sim0.5$–$1.0$ dex in $\dot{M}$, and $50$–$150$ km/s in $v_\infty$. Line-force parameters are uncertain by $\sim0.1$–$0.2$.
}
\end{table*}
\end{appendix}

\end{document}